# Angle-tunable intersubband photoabsorption and enhanced photobleaching in twisted bilayer graphene

Eva A. A. Pogna[1], Xianchong Miao[2], Driele von Dreifus[3], Thonimar V. Alencar[4], Marcus V. O. Moutinho[5], Pedro Venezuela[6], Cristian Manzoni[7], Minbiao Ji[2], Giulio Cerullo[7], Ana Maria de Paula[3]

[1] *Istituto di Nanoscienze CNR-NANO, Laboratory NEST, Piazza San Silvestro 12, Pisa 56127, Italy*
[2] *Laboratory of Surface Physics and Department of Physics, Fudan University, Shanghai 200433, China*
[3] *Departamento de Física, Instituto de Ciências Exatas, Universidade Federal de Minas Gerais, Belo Horizonte-MG 31270-901, Brazil*
[4] *Departamento de Física, Instituto de Ciências Exatas e Biológicas, Universidade Federal de Ouro Preto, Ouro Preto-MG 35400-000, Brazil*
[5] *Núcleo Multidisciplinar de Pesquisas em Computação - NUMPEX-COMP, Campus Duque de Caxias, Universidade Federal do Rio de Janeiro, Duque de Caxias- RJ 25265-970, Brazil*
[6] *Instituto de Física, Universidade Federal Fluminense, UFF, Niterói-RJ 24210-346, Brazil*
[7] *IFN-CNR, Dipartimento di Fisica, Politecnico di Milano, Piazza L. da Vinci 32, Milano 20133, Italy*

## Abstract

Van der Waals heterostructures obtained by artificially stacking two-dimensional crystals represent the frontier of material engineering, demonstrating properties superior to those of the starting materials. Fine control of the interlayer twist angle has opened new possibilities for tailoring the optoelectronic properties of these heterostructures. Twisted bilayer graphene with a strong interlayer coupling is a prototype of twisted heterostructure inheriting the intriguing electronic properties of graphene. Understanding the effects of the twist angle on its out-of-equilibrium optical properties is crucial for devising optoelectronic applications. With this aim, we here combine excitation-resolved hot photoluminescence with femtosecond transient absorption microscopy. The hot charge carrier distribution induced by photo-excitation results in peaked absorption bleaching and photo-induced absorption bands, both with pronounced twist angle dependence. Theoretical simulations of the electronic band structure and of the joint density of states enable to assign these bands to the blocking of interband transitions at the van Hove singularities and to photo-activated intersubband transitions. The tens of picoseconds relaxation dynamics of the observed bands is attributed to the angle-dependence of electron and phonon heat capacities of twisted bilayer graphene.

Van der Waals heterostructures composed of stacked single-layer crystals represent a promising platform for novel electronic, photonic and spintronic devices [1-8]. Wide tunability of the optoelectronic properties is achieved by both selection of the stacked materials and control of the relative rotation between the crystal registries [9, 5,10-12], i.e. the twist angle. In the case of two layers of graphene, strong interlayer coupling drives the hybridization of the electronic states and can lead to the formation of moiré super-lattices for finite twist angle. The rotational misalignment of twisted bilayer graphene (tBLG) enables to explore also new properties as topological valley transport [13], superconductivity [14] at the magic twist angle $θ \sim 1.1°$, or the topologically protected 1D chiral states appearing at the moiré domain walls [15]. The Dirac cones associated to the two layers in tBLG are no longer aligned in the energy- momentum landscape and two saddle points appear at their low-energy intersections. The saddle points contribute to van Hove singularities (vHs) in the density of states, which occur at energies $E_{vHs}$ that depend on the twist angle [16]. The vHs are detected as a peak in the optical conductivity of tBLG [17-19] and they have been related to the increase of hot photoluminescence (PL) for resonant excitation [20]. These vHs appear at much lower energies than in single-layer graphene (SLG) [16,21-24] and allow angle controlled visible and near-infrared (NIR) absorption enhancement. This property has been exploited in a tBLG based photo-detector to selectively increase the photocurrent generation at the vHs [25].

The changes in crystal symmetry, electronic band structure and optical conductivity come along with modifications of the relaxation pathways available to charge carriers driven out of equilibrium. The carrier dynamics in SLG is well-known as it has been intensively studied by means of ultrafast transient absorption [26-31] (TA), time-resolved photocurrent [32,33] and photoemission spectroscopies [34-36]. Excitation with femtosecond light pulses produces out-of-equilibrium distributions of electrons and holes in the conduction and valence bands, which thermalize in ~ 50 fs via Coulomb two-body scattering to hot Fermi-Dirac (FD) distributions [30, 31]. The relaxation of the hot carriers proceeds through scattering with strongly coupled optical phonons (SCOP) till a common temperature is reached in about 200 fs. The cooling of the thermalized electron-phonon bath is finally driven by the lattice and achieved in less than 10 picoseconds via anharmonic decay of the hot phonons. In defected SLG, hot carriers can release their excess energy also by defect-mediated emission of acoustic phonons (supercollision mechanism), which occurs on a ps-time scale dependent on substrate and defect densities [31,33,37-43]. The hot carrier distribution inhibits interband absorption over a broad energy range due to Pauli blocking, so that in the TA spectra of SLG a decreased absorption, or photo-bleaching (PB), signal is detected.

In Bernal stacked bilayer graphene, together with the PB band, *Limmer et al.* [44] observed a photoinduced absorption (PA) with a single decay time of 5 ps, about twice as long as the PB signal. The PA has been attributed to transitions between subbands formed within the conduction and the valence bands, named intersubband (ISB) transitions, which are absent in SLG since it lacks a subband structure. Also time- and angle-resolved photoemission spectroscopy (TARPES) has demonstrated long-living (up to 10 ps) population inversion in bilayer graphene [45,46] attributed to the presence of a small gap. Previous studies in tBLG by TA microscopy [47,48] and TARPES [49] have explored the twist-angle dependence and shown long decay times attributed to bound excitons and to states reached by two-photon absorption [48, 50].

Here we study the out-of-equilibrium optical properties of tBLG as a function of the twist angle by femtosecond TA microscopy with high sensitivity, which allows to explore low excitation regimes in which angle-dependent contribution from ISB transitions emerges. Specifically, we investigate heterostructures with the vHs located in the NIR and twist angle $\theta$ in the range of ~ 6°-8°. The tBLG samples are prepared by chemical vapor deposition as described in the Methods and they have tens of μm linear dimension. A lattice rotation in the real space by an angle $\theta$ corresponds to a rotation of $\theta$ in the phase-space of the hexagonal Brillouin zones (BZs), as sketched in Fig. 1(a-b). Accordingly, the Dirac points of the two layers (see Figure 1(c)), $K_A$ and $K_B$, no longer coincide and are displaced by a momentum *ΔK = 2K sin(θ/2)*, where *K = 4π/3a* and *a* ≈ 0.246 nm is the SLG lattice parameter [51]. The interlayer interactions perturb the Dirac cones of each layer causing the formation of minigaps with the same energy in the conduction and valence bands at the vHs, as is schematically shown in Fig. 1(c).

Figure 1(d) shows the optical microscope image of one of the investigated samples, named stack 1, measured in reflection. It consists of a large single-crystal SLG, with an inner brighter region, where a second graphene layer is stacked to form a bilayer single crystal. A first determination of the mutual orientation of the lattices in the bilayer domain is obtained by extrapolation of the hexagonal habits of the outer/inner growth regions, corresponding respectively to the 1st/2nd layer [52]. In stack 1, the hexagons are rotated by an angle $\theta = 7°\pm 1°$. The angle determination from the optical images is affected by a large uncertainty related to the definition of the single-crystals shape. Twist angle is usually measured by field high-resolution transmission electron microscopy [53,54], scanning tunneling microscopy [16] and lattice resolution atomic force microscopy [55]. Here we adopt an all-optical, far-field approach, based on measuring the hot-photoluminescence excitation (hot-PLE) spectrum under femtosecond laser excitation. As previously observed [20], due to the enhancement of pump absorption, a peak is expected in the bilayer emission when the laser is tuned at the vHs. In Fig. 1(e) the hot PL intensity of tBLG normalized to that of SLG ($I_{tBLG}/I_{SLG}$) is shown as function of laser excitation energy. The experimental data are fitted with a Gaussian function centered at $E_{vHs}$ = 1.3880 ± 0.007 eV. From the energy of the vHs, we extract a more accurate estimation of the twist angle considering the correlation $\theta\ [°] = A - \sqrt{(B - C\ E_{vHs}\ [eV])}$, with $A$ = 45.380, $B$ = 2.030 × $10^3$, $C$ = 4.186 × $10^2$ extracted from analysis of the calculated joint density of states (JDOS) as a function of $\theta$ reported in the Fig. S-1 of the Supplementary Material (SM).

We obtain the value $\theta$ = 7.31° ± 0.04° for stack 1, in good agreement with the evaluation from the optical image inspection. The hot PL map in Fig. 1(e) for laser excitation tuned at 1.4 eV, close to $E_{vHs}$, reports indeed a ≈ 2 times higher emission intensity at the tBLG compared to the SLG, uniform across the bilayer domain.

Extra peaks have been observed in visible, IR and UV Raman spectra of tBLG compared to SLG. Accordingly, we characterize the heterostructure by micro-Raman spectroscopy using a visible laser tuned at 633 nm. Figure 1(f) shows the Raman spectra at the SLG (red line) and tBLG (blue line) regions. The G-band peak at 1584 $cm^{-1}$, the 2D peak at 2648 $cm^{-1}$ and an additional peak at 1621 $cm^{-1}$ are observed for tBLG, while for SLG these band are observed at 1588 $cm^{-1}$ (G) and 2645 $cm^{-1}$ (2D).

In the bilayer, the G-band peak has a two times higher intensity, as shown in the inset of Fig. 1(f). The enhancement of the G-band and the extra peak, $L_a$, testify the formation of a new band structure by interlayer coupling. It is thus an indication of the high crystalline quality of the samples and of a clean interface between the two monolayers forming the tBLG. The $L_a$ peak above the G-band position for the bilayer is the mode activated by intralayer electron-phonon scattering process from the unfolded longitudinal optical phonon (LO) [56] branch. The measured $L_a$ frequency of 1621 $cm^{-1}$ is consistent with the theoretical predictions [56] for a tBLG with $\theta \sim 7°$. The LO dispersion is expected to have a non-monotonic dependence on the twist angle, with a maximum frequency at $\theta \approx 10°$ but only slowly varying in the range of interest, thus making it unpractical to precisely define the twist angle only from the Raman data.

We apply ultrafast TA microscopy to study the photophysics of tBLG samples as function of the twist angle. A NIR pump pulse with photon energy $E_{pump}$ is used to photo-excite the sample off-resonance with the vHs ($E_{pump}$

= 0.8 eV stack 1 and $E_{pump}$ = 1.19 eV for stacks 2-3), while a second tunable pulse, the probe, monitors the relaxation of the induced hot carriers' distributions as function of pump-probe time delay *t*. Pump and probe beams are collinearly coupled to a confocal microscope to achieve a ≈ 1 µm spatial resolution. The differential transmission (*ΔT/T*) is measured as a function of the probe photon energy $E_{probe}$, using two distinct high sensitivity TA microscopes described in the Methods, guaranteeing a spectral coverage from 0.85 eV to 1.65 eV. Selected *ΔT/T* maps at fixed $E_{probe}$ and time delay *t* = 150 fs are reported in Figure 2, for three tBLG stacks with different *θ*. As expected, the SLG regions exhibit a PB signal at all the investigated $E_{probe}$, detected as a positive *ΔT/T* signal. The PB signal only shows a smooth monotone increase towards lower energy, following the momentum-spreading of the hot-carriers FD distribution.

This behavior contrasts to that of tBLG, where a complex dependence on $E_{probe}$ emerges showing also a PA signal. Specifically, under low photo-excitation fluences, we identify two characteristic components in the transient response: a clear peak in the PB and the appearance of PA at lower energy. The PB peak corresponds to a factor 4-6 enhancement with respect to the SLG signal and it is centered in proximity to $E_{vHs}$ (see first column of maps of Figure 2). The PA is detected for $E_{probe}$ < $E_{vHs}$. At lower probe energies, about 300 meV below $E_{vHs}$, the *ΔT/T* signal of tBLG becomes barely distinguishable from that of SLG. While a similar behavior is observed in all the investigated samples, the energy position of PB and PA bands changes with *θ*. The larger is the twist angle, the higher is $E_{probe}$ at which similar tBLG/SLG contrast is observed. In stacks 2 and 3, bilayer regions with different twist angle are present, so that the signal intensity evolves differently according to their orientation.

For all samples, the precise value of the twist angle is determined by recording the hot-PLE spectrum. The experimental *ΔT/T* spectra of monolayer and bilayer regions at *t* = 150 fs for several tBLG samples are reported in Figure 3(a), together with the PLE spectra used to extract $E_{vHs}$ and *θ* (blue symbols).

As *θ* decreases, the overall TA spectrum experiences a red-shift. The PLE peak almost overlaps with the PB peak, with the larger mismatch appearing in the stack with the most intense PA band. The signal near the vHs, indeed, results from the interplay of PA and PB contributions whose relative intensity varies with the twist angle and excitation fluence. Remarkably, the PA emerges at low photo-excited carrier densities of the order of *n* = 1 × $10^{13}$ cm$^{-2}$, as estimated from an incident fluence of 50 µJ cm$^{-2}$ considering 4% absorption. At an incident fluence of 350 µJ cm$^{-2}$, corresponding to *n* = 7 × $10^{13}$ cm$^{-2}$, the PA band disappears and a PB is observed over all the spectral range (see Fig. S-2 and S-3 of SM). The lack of previous evidences [47,48] of contribution from the ISB to the TA of tBLG could be traced back to the higher absorbed photon flux attained by pumping resonantly to the vHs. This dominance of the PB over the PA for high laser fluence was also observed for the intra- and inter-band transitions in suspended single layer graphene [31].

The dependence of the experimental TA spectrum on crystal orientation is traced back to the modifications of the phonon modes, electronic structure and of the JDOS, which we theoretically compute as a function of the twist angle and the hot-electron temperature $T_e$ (see details in Methods section) [57]. Figure 3(b) shows the calculated JDOS considering the interband and intersubband transitions between electronic states for the initial hot carrier temperature of 1500 K and same twist angles as in panel (a). The electronic states populated by the pump pulse allow for the ISB transitions. The temperature of the quasi-equilibrium hot electron distribution is obtained considering a three temperature model [58] (see Fig. S-4 and S-5 and details in the SM). The black (red) curves in Fig. 3(b) show the JDOS for the interlayer ISB (interband-IB) transitions as a function of the twist angle $\theta$. The ISB curves are calculated for the electron and hole transitions as indicated in Fig. 1(c) by the solid and dashed red arrows, respectively. The higher energy peak in the JDOS for ISB transitions corresponds to holes' excitations, while the lower-energy peak to the electrons. The difference in energy for the electron and hole ISB transitions is due to the small difference of their curvatures. Note that for all the twist angles, the electron ISB peak appears about 50 meV below the IB peak due to the vHs transitions, while the hole ISB peak is almost overlapped with the vHs transitions (see Fig. S-1 of SM). Figure 3(c) shows a comparison of both IB and ISB JDOS for carrier temperature of 1500 K and room temperature (300 K); the blue arrows indicate that the ISB JDOS intensity increases with the temperature while the IB JDOS intensity decreases. The intensity of the JDOS is related to the dynamical conductivity (or absorption). In this way, the difference of the JDOS at these two temperatures is proportional to the $\Delta T/T$ spectrum and qualitatively explains the opposite sign of the contributions from ISB and IB transitions to the TA signal.

The increase of the JDOS associated to ISB transitions due to photo-excitation corresponds to the PA signal observed at energies just below $E_{vHs}$ in the experimental data of Figure 3(a). The ISB peak for holes is very close to the $E_{vHs}$ transition that dominates the $\Delta T/T$ signal at these energies. The strong angle dependence of the PA energy position results in ~ 250 meV tuning with only a ~ 1.26° twist angle change. Model predictions of the PA energy position based on the JDOS calculation, with the folding method, agree very well with the experimental data. In the SM, the expected values for a large range of twist angles (from 0° to 30°) are reported in Fig. S-1 and appear always slightly red-shifted with respect to the vHs transition.

Figure 4 reports the $\Delta T/T$ dynamics of stack 1. A similar relaxation scenario is observed for all the stacks, once the shift of energy position of the transient spectrum is taken into account (see Fig. S-6 in the SM). The recovery dynamics of tBLG is consistently different from that of SLG. The dynamics of tBLG (full colored dots) is compared to that of SLG (open black symbols) for $E_{probe}$ corresponding to the main features of the optical response: the PB peak at the vHs, the PA corresponding to ISB transitions and the low-energy PB band. The $\Delta T/T$ dynamics in SLG is energy-independent and it is described by a bi-exponential decay convoluted to an instrumental response function (IRF) of ≈ 120 fs. The extracted time constants are $\tau_1^{SLG}$ = 120 ± 20 fs and $\tau_2^{SLG}$ = 1980 ± 60 fs. These values are characteristic of SLG samples with low defect density [43]. The transient transmissivity depends superlinearly on $T_e$ through the FD distribution function, so that the time evolution of $T_e$ and $\Delta T/T$ are not correlated in time. However, while the fast relaxation component is strongly influenced by the superlinearity, the slow relaxation component of transient transmissivity is generally assumed as a reliable estimation of cooling time via anharmonic coupling of the thermalized bath of hot-electron and SCOPs with the acoustic phonons [59]. The $\Delta T/T(t)$ dynamics of tBLG, on the contrary, strongly depends on the transition energy even though, away from the PA and the vHs, it is very similar to that of SLG. At the vHs, the PB not only has a much higher initial intensity than for SLG, but it decays on a longer time scale. The dynamics, after deconvolution with the IRF, is best fit by a biexponential decay with $\tau_1^{tBLG}$ = 140 ± 20 fs and $\tau_2^{SLG}$ = 2500 ± 100 fs. The PA band is best described by a three-exponential decay with time constants $\tau_1^{tBLG}$ = 120 ± 30 fs, $\tau_2^{tBLG}$ = 1960 ± 40 fs and $\tau_3^{tBLG}$ = 70 ± 2 ps with the additional decay component responsible for the tail extended to 100 ps in Figure 4(b). Such long lifetimes can be explained by the peculiar temperature dependence of the JDOS that emerges at finite twist angles. We model the decay curves by the ratio of the calculated JDOS as a function of the electron temperature $T_e$, see Fig. 4(c-d). We calculate the time evolution of $T_e$ from a three temperature model, considering the electron energy relaxation through strongly coupled and weakly coupled phonon modes [58]. The specific heats for the electrons, strongly coupled phonons and weakly coupled phonons are obtained considering their density of states from tight-binding [60,57] and force constant [61] methods parameterized by first principles calculations. For tBLG, it is important to include the additional low frequency and high density phonon modes at the Γ point, that are not present in SLG, as representing the out-of-plane breathing of adjacent layers interacting via weak van der Waals forces (details in the SM). The calculated *ΔT/T(t)* dynamics, both at the vHs transition and at the ISB transition, reproduce very well the experimental results. Indicating that the density of states enhancement at the vHs and the changes in the electron and phonon heat capacities can explain the observed long decay of the PA and PB bands.

To conclude, we demonstrate that at low excited carrier density the transient optical response of tBLG shows two peaks, one due to an photoinduced absorption from intersubband transitions and another one due to photobleaching of the interband transitions in resonance with the vHs ($E_{vHs}$). Similarly, to the vHs, the energy positions of these peaks can be controlled with the twist angle. At the vHs, we observe a carrier cooling time which is about 30% longer than for SLG, while the photoinduced intersubband signal shows a long-living tail up to 100 ps. A simple model, based on the JDOS enhancement and on the evolution of the electronic temperature as the hot electrons equilibrate with the tBLG phonon modes, explains well the experimental TA signal, both for the peaks position and for the time decays, without recurring to more complex multiparticle excitations as bound excitons or non-linear processes such as two-photon absorption.

**Methods**

***Sample preparation and characterization*** The twisted bilayer graphene single crystals were prepared by ambient pressure chemical vapor deposition (AP-CVD) synthesis on polycrystalline Cu foils. The nucleation conditions on Cu foils are controlled to allow both constituent layers to form as hexagonal single crystals, and each tBLG grain has a specific twist angle. The samples were transferred to a transparent amorphous quartz substrate for the optical images and spectroscopy measurements. More details about the sample preparation are presented in Ref. [62].

To select the samples for TA studies we first measured the sample twist angle by an analysis of the flake shapes on optical microscopy white light images. The optical images were obtained using an Olympus microscope (DP72) in reflection mode with a 100X objective lens. To obtain the twist angle we drew a hexagon for each layer and measured the angle between the edges of the two hexagons [52], as can be seen for the sample in Figure 1(d). We also characterized the sample twist angle by Raman spectroscopy using an alpha300R WITec confocal Raman spectrometer at excitation energy of 1.96 eV (633 nm). The Raman spectra were measured in backscattering geometry on a triple monochromator spectrometer equipped with a liquid-N2 CCD. All measurements were performed at room temperature.

For the PLE measurements a PL image is acquired for each excitation laser energy, that is tuned to cover the range of the vHs transition for each sample. The PL intensity is averaged at the bilayer region and divided by the PL at the monolayer region to obtain each point in the PLE spectra. The PL signal is measured in back scattering, separated from the excitation laser by a dichroic mirror (Semrock FF665-Di02). A band pass filter (75 nm centered at 610 nm) is used in front of the detector to further block any scattered laser light (Chroma 610/75M).

***High-sensitivity TA microscopy*** The TA data here reported have been acquired using two distinct microscopes working in complementary probe energy ranges, enabling an overall spectral coverage of 1 eV. The microscope used for the probe energy range of 0.8-1.4 eV is based on a Er-doped fiber laser (Toptica-Femtofiber pro) generating 300 mW, 150 fs pulses centered at 1550 nm with 40 MHz repetition rate. A portion of the output of the laser is used as the pump pulse tuned to 0.8 eV and is modulated with an acousto-optic modulator operating at 1 MHz. The probe pulse extending from 0.73 to 1.45 eV is obtained by super continuum (SC) generation from the laser fundamental focused in a Er-doped highly-nonlinear fiber. The low-energy branch of the SC and the fundamental are filtered out with a short pass filter cutting at 1500 nm. The pump and the probe beams are collinearly focused on the sample with an objective (Olympus-LCPLN-IR with magnification 100 X and NA = 0.85) to a spot size of about 1 μm. The probe transmitted by the sample is collected with an 8 mm focal length achromatic doublet, spectrally filtered with a monochromator with 5 nm resolution and detected by an InGaAs amplified photodiode with 4 MHz bandwidth. The TA signal is measured with a lock-in amplifier with 300 ms effective time constant, resulting in a *ΔT/T* sensitivity of $10^{-7}$. TA dynamics is monitored by changing the pump-probe delay with an optical delay line, while the TA maps (images) at fixed probe photon energy and time delay are acquired by moving the sample with a motorized three-axis piezo-stack linear stage (Newport NPXYZ100). The microscope used for the probe energy range of 1.25-1.65 eV is based on a commercial femtosecond optical parametric oscillator (OPO, Insight DS+, Newport, CA) used as the laser source, with the fundamental 1040 nm (1.1923 eV) beam as the pump and the tunable output 750-990 nm (1.252-1.653 eV) as the probe. The pump beam is modulated at 20 MHz by an electro-optical modulator, collinearly combined with the probe beam and delivered into a laser-scanning microscope (FV1200, Olympus). The laser beams are focused onto the sample with an air objective (UPLSAPO 20X, NA = 0.75, Olympus) to about 1.5 μm diameter, raster scanned by a pair of galvo mirrors, transmitted through the sample, passed through an optical filter to block the pump beam, and then directed onto a large-area silicon photodiode. Pump fluence is kept at approximately 50 μJ cm$^{-2}$ (2 x $10^{14}$ incident photons/cm$^2$). The pump and probe beams are combined with a dichroic mirror (DMSP1000, Thorlabs) and aligned collinearly. The probe beam is optically filtered by two short-pass filters (FF011010/SP-25, Semrock), collected by a photodiode, and demodulated with a commercial lock-in amplifier (HF2LI, Zurich Instruments) to extract the TA signal (*ΔA*). A pixel dwell time of 2 μs and image size of 320×320 pixels are used in the experiments. The cross-correlation between pump and probe pulses at the sample is measured to be ~ 0.3 ps, and the time delay interval between adjacent images is set to *Δt* = 133 fs. A single *t* scan of 20 ps takes about 1.5 min, and a complete spectral scan takes about 30 min.

***Theoretical model*** The electronic structures were obtained by folding the SLG calculation results. For SLG, we follow the procedure given in Ref. [60, 57], where the electronic structure calculations are based on a fifth neighbors tight-binding approach, with one orthonormalized $p_z$ orbital per site, in which the hopping parameters are fitted to reproduce density functional theory (DFT) calculations with many-body corrections.

The electronic transitions between bands in tBLG are captured considering two SLG BZs twisted by a certain angle *θ*. In this way, we considered only four electronic bands for the tBLG, one valence and one conduction band for each layer. For low temperatures, only the electronic IB transitions are possible. The magenta arrow in Fig. 1(c) shows the IB that coincides with the transition between the vHs, that gives rise to a peak in the JDOS at an energy $E_{vHs}$ which is a function of *θ*. When the temperature increases, ISB transitions between valence, or between conduction bands, are also possible (red arrows in Fig. 1(c)).

The allowed IB and ISB transitions do not depend on the size of the minigap, which appears due to the hybridization of the eigenstates near to the band intersections. Our model cannot reproduce the interaction between layers and the atomic relaxation is not taken into account. However, *Havener et al.* [19] have shown that, although the interlayer interactions perturb the bands from each layer producing minigaps, which splits the valence and the conduction bands in tBLG, the difference in energy between the minigaps is generally too small to be seen directly in the PL and absorption spectra, contributing only to the peak broadening. Also, since the allowed electronic transitions do not depend on the size of the minigap and the PL peak is usually symmetric, which means that both transitions from minigaps to the conduction and the valence bands contribute equally, the energy position of the JDOS peak obtained by the folding approach is supposed to give a good estimation of the twisted angle. Therefore, the folding approach used in our simulations, which does not consider interaction between layers, is supposed to give good results [19, 56].

In our model, it is possible to analyze the JDOS for IB and ISB transitions separately, as a function of the electronic temperature. We compute the JDOS as

$$J(T_e, E) = \frac{\gamma}{\pi N_k} \sum_{i,j,k} \frac{f(E_{i,k}, T_e) - f(E_{j,k}, T_e)}{(E_{i,k} - E_{j,k} + E)^2 + \gamma^2} \quad (1)$$

where the sum is performed on a uniform grid of $N_k$ = 2400 × 2400 k-points over the two dimensional BZ and for the different pairs of bands *i*; *j* while $\gamma$ =1 × $10^{-2}$ eV is a small energy broadening parameter related to the inverse of the electron-hole lifetime. The function $f(E_{i,k}, T_e) = [1 + \exp(E_{i,k}/T_e)]^{-1}$ is the Fermi function for a specific band energy *i* of the state ***k*** and temperature $T_e$ (the chemical potential *μ* is close to the Fermi level and both are set to zero).

Using only IB transitions in the JDOS calculation we find the energy peaks corresponding to the $E_{vHs}$ for a specific twist angle *θ*. For the same values of $E_{vHs}$ obtained by the PLE peaks, we compute the normalized JDOS for IB and ISB transitions, as a function of the probe energy for $T_e$ = 1500 K, which is the estimated temperature for the hot electrons excited by the pump laser (see Fig. S-4 and S-5 in the SM). We see that the peaks corresponding to ISB transitions are always at lower energies than the IB transitions between the vHs, as shown in Fig. 3(c). An estimation of the *ΔT/T* dynamics can be obtained assuming that electron-radiation matrix element is constant for all transitions. In this way, the dynamical conductivity σ (or absorption α) becomes proportional to the JDOS which we compute including both the IB and the ISB as function of the $T_e$. Details of the $T_e$ time-evolution calculation are described in the SM.

## Acknowledgements

We acknowledge financial support from Graphene FET Flagship Core 3 Project, Grant No. 881603 and the Brazilian funding agencies: FAPERJ (grant number E-26/010.101126/2018), Fapemig, CNPq, Capes and INCT Carbon Nanomaterials. Prof. Po-Wen Chiu is gratefully acknowledged for preparing the tBLG samples.

**Supplementary Material**: twist angle estimation from $E_{vHs}$, fluence dependence of TA spectra, electron temperature calculation, relaxation dynamics of tBLG stacks) is available.

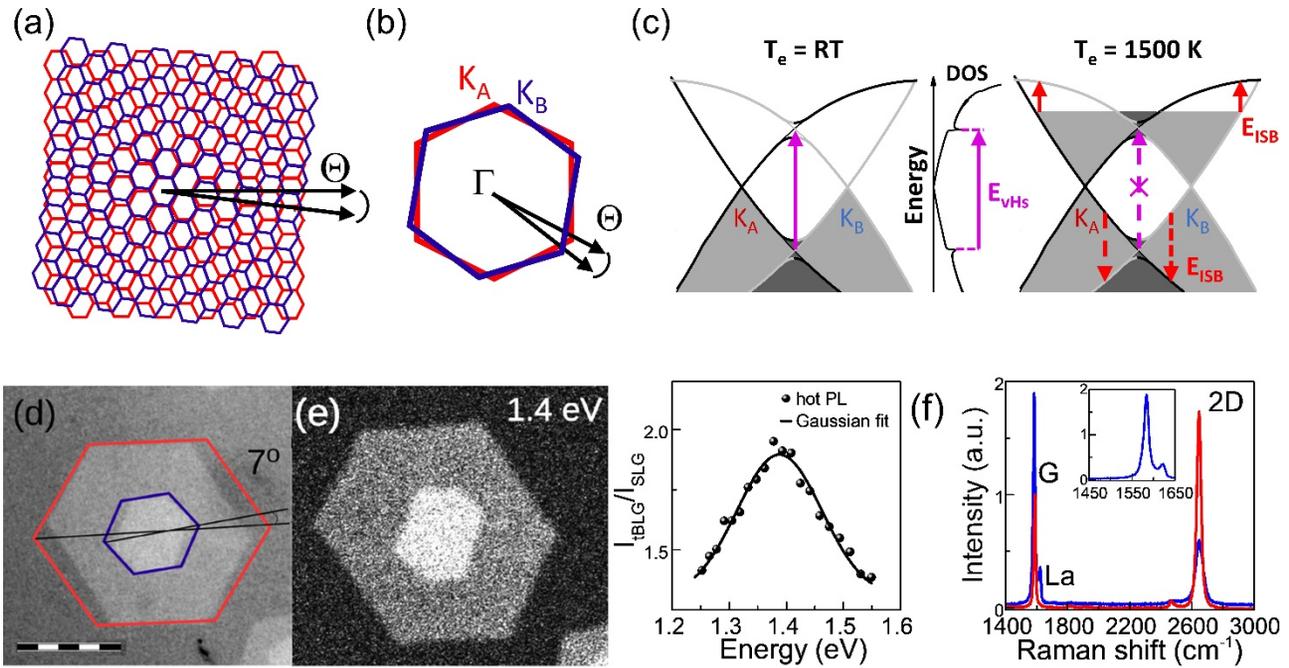

**Figure 1** (a) Sketch of the crystal structure of a tBLG with twist angle $\theta$; (b) the corresponding BZs of two rotated graphene layers as the red and blue hexagons; (c) energy structure indicating the optical transition at the vHs at photon energy $E_{vHs}$ (magenta line) and intersubband transitions from photo-excited FD of electrons (red solid lines) and holes (red dashed lines); (d) bright field optical image of stack 1 with blue and red hexagons identifying a layer rotation angle of about 7°, scale bar 10 μm. (e) PL intensity map for laser excitation energy of 1.4 eV, in resonance with the $E_{vHs}$ identified by the peak in the PLE intensity of tBLG ($I_{tBLG}$) divided by that of the corresponding SLG ($I_{SLG}$); (f) Raman spectra of stack 1 at the bilayer (blue line) and monolayer (red line) regions normalized by the intensity of the G peak. The inset shows a zoom-in of the bilayer G and $L_a$ peaks divided by the monolayer G peak.

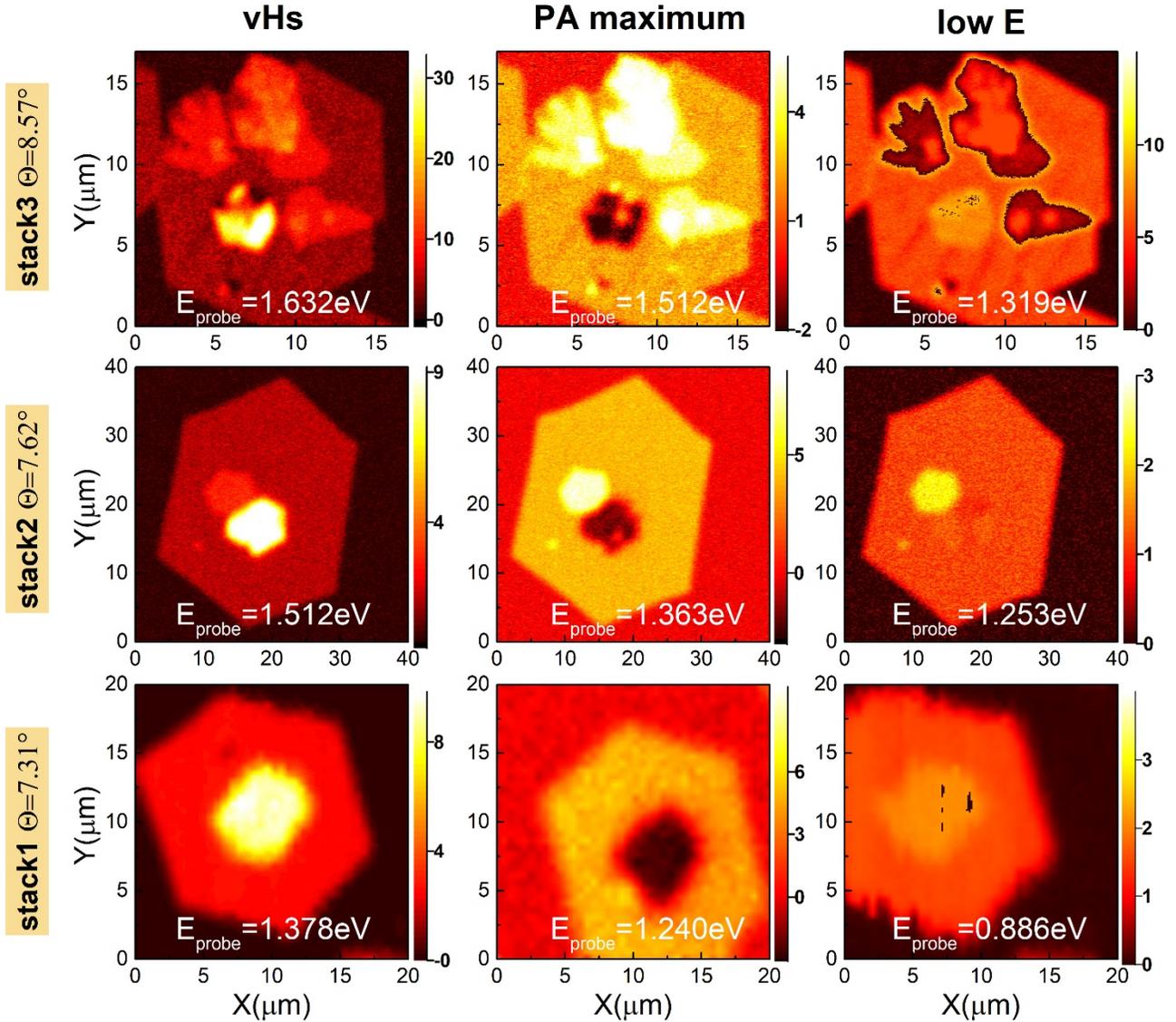

**Figure 2** Color map images of the transient transmissivity ΔT/T ($10^{-4}$) at time delay $t$ = 0, for three tBLG stacks with twist angle increasing from the lower to the upper row of panels: $\theta$ = 7.31°, 7.62° and 8.57°. The probe energies, for the first column of maps are taken in resonance with the vHs at $E_{probe}$=1.632, 1.512, 1.378 eV; second column at the maximum of the PA band at $E_{probe}$= 1.512, 1.363, 1.240 eV; third column with low-energy PB band at $E_{probe}$=1.319, 1.253, 0.886 eV, as indicated in the panels.

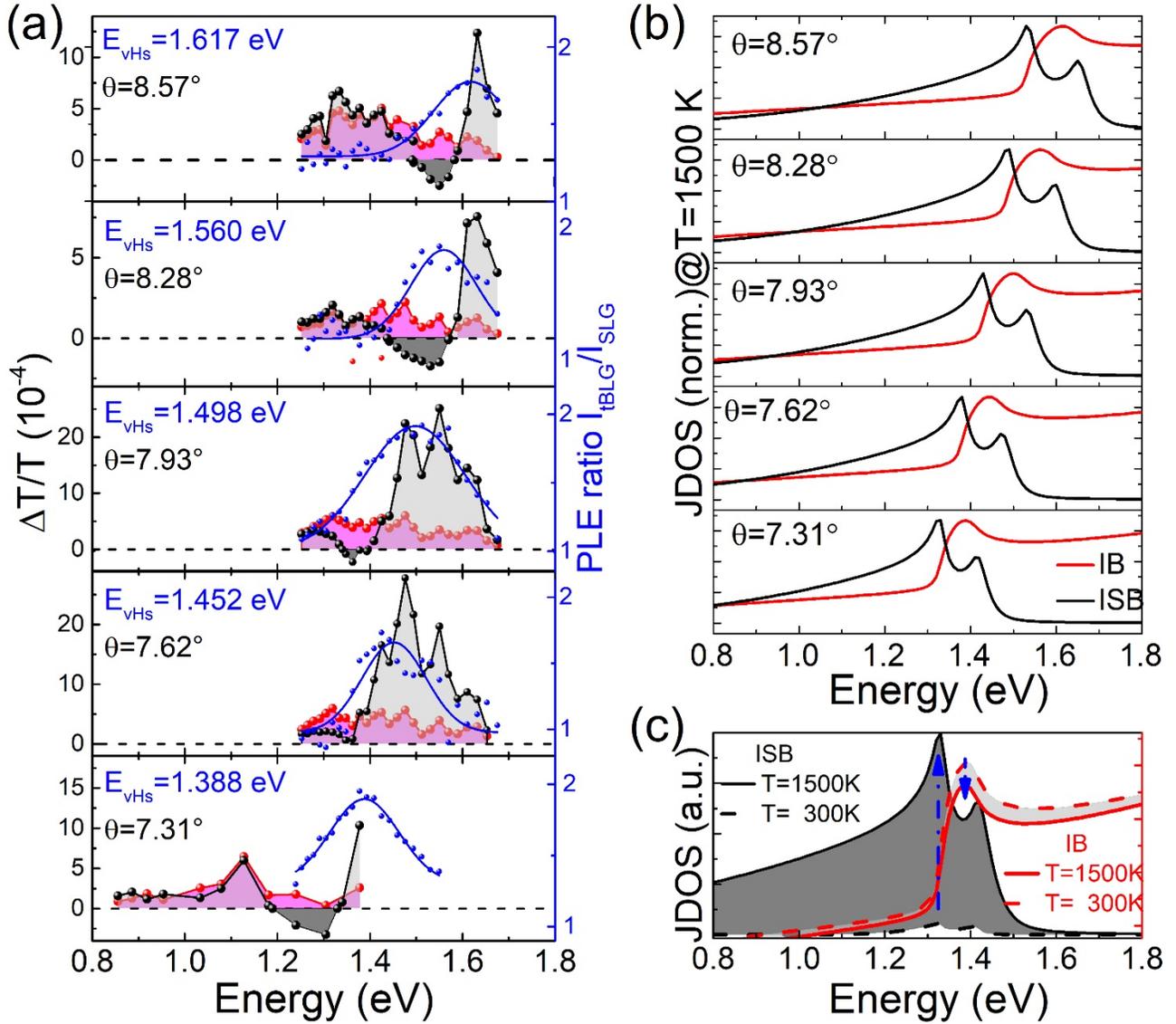

**Figure 3** Transient transmissivity spectra of SLG (red-dot and solid line) and tBLG (blackdot and solid line) at fixed time delay $t$ = 150 fs for samples with different $\theta$ as indicated by the labels. The blue dots are the PLE intensity ratio of the tBLG to the SLG fitted with a Gaussian function (blue solid line) peaked at $E_{vHs}$. (b) calculated JDOS for the ISB transition (black) and the $E_{vHs}$ transition (red) for a hot electron temperature of 1500 K. (c) A comparison of the IB and ISB JDOS for 1500 K and the room temperature 300 K. The difference of the JDOS at these two temperature is proportional to the $\Delta T/T$ spectrum.

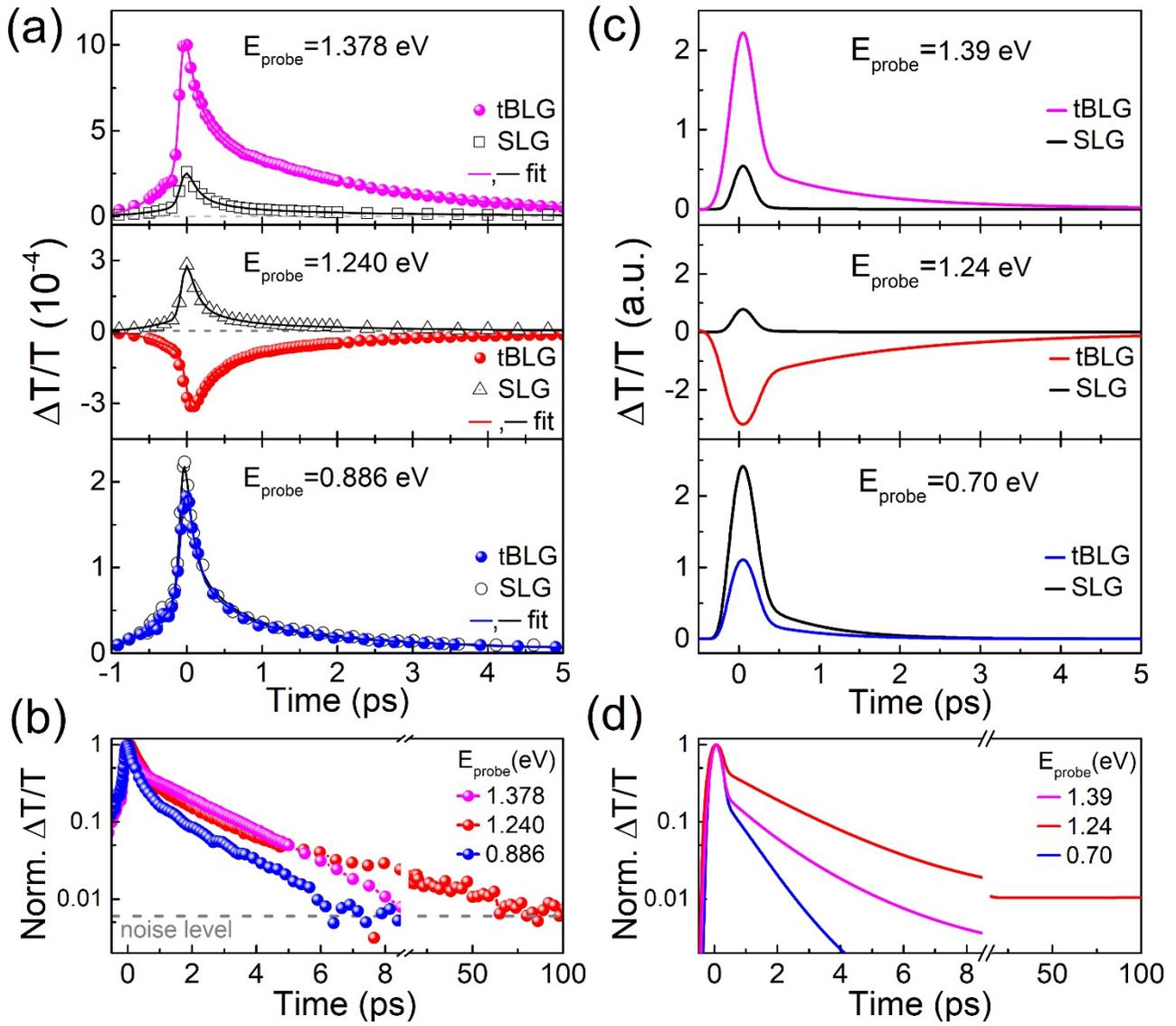

**Figure 4** (a) Transient transmissivity *ΔT/T* dynamics of tBLG (full colored dots) and SLG (open black symbols) for the stack with θ = 7.31° at probe energy $E_{probe}$ = 1.378 eV corresponding to $E_{vHs}$; 1.240 eV corresponding to the ISB; 0.866 eV in the low-energy PB band E < $E_{ISB}$. (b) Comparison of *ΔT/T(t)* normalised to the maximum and with an extended time scale. (c) Theoretical estimated decay curves based on the electron temperature time decay. In the simulations we use $E_{probe}$ = 1.39, 1.24 and 0.70 eV to best reproduce the experimental decay behavior. (d) Normalized theoretical curves obtained in (c).

# Supplementary Material

## 1. Twist angle estimation from $E_{vHs}$

The electronic band structure and the corresponding density of states are evaluated for different twist angles. The energy position of the calculated vHs and ISB electronic transitions for holes and electrons are reported in Fig. S-1 as function of the twist angle.

In analogy with Ref. 1 we find a convenient expression of $\theta$ fitting the theoretical data in the inset in the range 6.8° < $\theta$ < 8.6° with the function $\theta\,[°] = A - \sqrt{(B - C\,E_{vHs}[eV])}$, with A = 45.380, B = 2.030 × 10³, C = 4.186 × 10².

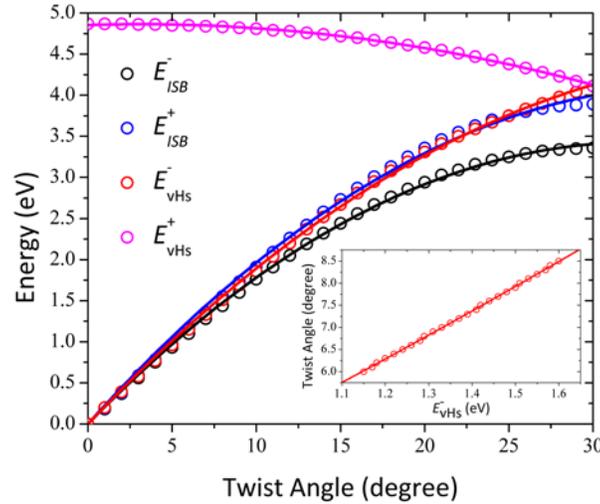

**Figure S-1** Calculated energy of vHs for electrons $E^{-}_{vHs}$ (red dots and line) and holes $E^{+}_{vHs}$ (magenta dots and lines) and of ISB transitions for electrons $E^{-}_{ISB}$ (black dots and lines) and holes $E^{+}_{ISB}$ (blue dots and lines) as a function of the twist angle. The inset shows for the vHs energy in our experimental region of twist angles together with the best fit function (red solid line).

To calculate the uncertainty in the angle estimation we derive the expression: $\delta\theta[°] = \frac{C}{2\sqrt{B - C\,E_{vHs}\,[eV]}}\delta E$, considering the $\delta E$ as the PLE peak width. From the PLE results, the fitted Gaussian widths give $\delta E \leq 0.01$ eV for all the stacks. For the bilayer stack with $E_{vHs}$ = 1.388 ± 0.007 eV, for example, we find that θ= 7.312° ± 0.04°.

## 2. Fluence dependence of transient absorption spectra

The intersubband (ISB) transitions energetically overlap with interband optical transitions. The hot-electron distribution causes simultaneously the bleaching of interband absorption due to Pauli blocking and the activation of ISB absorption with contributions to the transient transmissivity of opposite signs. The interplay of these contributions depends on the fluence and, at high density of photo-excited carriers, we observe that the bleaching is dominating over the photo-induced absorption over the entire probe photon energy range.

Accordingly, while the TA of SLG scales linearly with fluence, the spectra of tBLG change shape with fluence, as reported in Fig. S-2(a) for four different tBLG samples. The dynamics as function of pump-probe delay and probe wavelength is reported in Fig. S-2(b) at the two compared fluences for the stack with $\theta$ = 7.93°, taken as prototype of tBLG.

While the photobleaching signal increases with excitation power, due to the increase in initial electron temperature and to the superlinear dependence of $\Delta T/T$ on $T_e$, the instantaneous PA signal decreases. In Fig. S-3 we report the pump power dependence of relaxation dynamics of the stack with $\theta = 7.31°$ at probe energy $E_{probe} = 1.240$ eV corresponding to the ISB, and $E_{probe} = 0.956$ eV in the low-energy PB band.

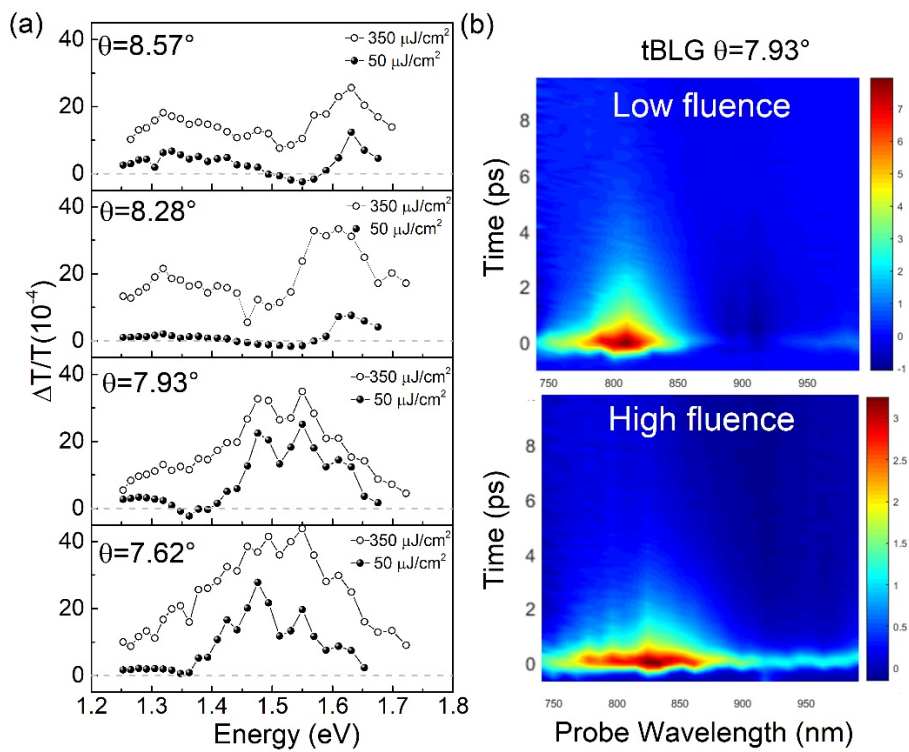

**Figure S-2** (a) $\Delta T/T$ spectra of tBLG at incident fluence of 350 µJ cm$^{-2}$ (open symbols) and 50 µJ cm$^{-2}$ (full symbols) for different twist angles $\theta$; (b) $\Delta T/T$ as a function of probe wavelength and time delay at low incident fluence 50 µJ cm$^{-2}$ (upper panel) and high incident fluence 350 µJ cm$^{-2}$ (lower panel) for the tBLG with $\theta = 7.93°$.

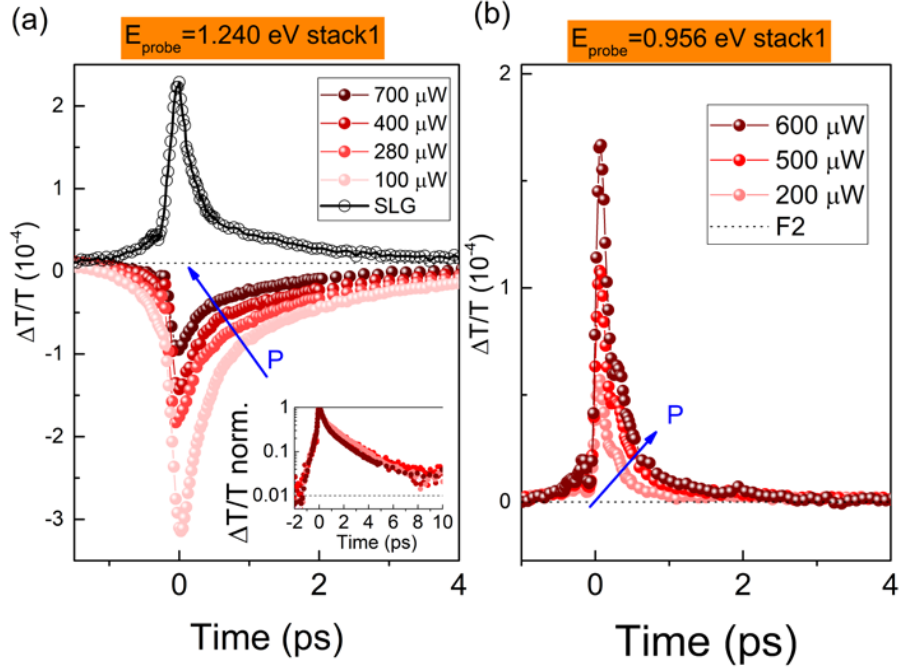

**Figure S-3** Transient transmissivity dynamics at different excitation power at probe photon energy (a) $E_{probe}$ = 1.240 eV corresponding to the ISB and (b) $E_{probe}$ = 0.958 eV in the low-energy PB band. Normalised dynamics at $E_{probe}$ = 1.240 eV are reported in the inset of panel a.

### 3. Calculating the electron temperature

In order to obtain the electron and phonon temperatures as a function of time we adopt a three temperature model [2]. This model considers that after a transition to an out-of-equilibrium state the electrons rapidly exchange energy with phonons via electron-phonon interaction achieving a quasi-equilibrium Fermi distribution. Considering the occupation number of electrons and phonons as Fermi-Dirac and Bose-Einstein distributions respectively, we can associate an effective temperature for the electrons, $T_e$, and for phonons, $T_v$ which are separated into strongly coupled (v = sc) and weakly coupled (v = wc) phonons. Thus, a system of coupled equations for the three temperatures ($T_e$, $T_{sc}$, $T_{wc}$) is used to simulate the dynamics of the electrons and phonons after the laser pulse excitation.

For a Gaussian pulse $I(t)$ with fluence F and pulse duration $\tau_p$, the temporal dependence of the effective temperatures is given by:

$$\frac{dT_e}{dt} = \frac{I(t)}{\beta c_e} - \frac{g_{sc}}{c_e}(T_e - T_{sc}) - \frac{g_{wc}}{c_e}(T_e - T_{wc})$$

$$\frac{dT_{sc}}{dt} = \frac{g_{sc}}{c_{sc}}(T_e - T_{sc}) - \frac{1}{\tau_o}(T_{sc} - T_{wc})$$

$$\frac{dT_{wc}}{dt} = \frac{g_{wc}}{c_{wc}}(T_e - T_{wc}) + \frac{c_{sc}}{c_{wc}\tau_o}(T_{sc} - T_{wc})$$

(1)

where β is a dimensionless parameter which determines the fraction of the pulse energy absorbed by the electrons and $\tau_o$ = 2.5 ps is the optical phonon lifetime [3-4]. The respective specific heats for the electrons ($c_e$), strongly coupled phonons ($c_{sc}$) and weakly coupled phonons ($c_{wc}$) are obtained by the following integrals:

$$c_e = \int_{-\infty}^{\infty} \epsilon N(\epsilon) \frac{\partial f_D(\epsilon, T_e)}{\partial T_e} d\epsilon \qquad (2)$$

$$c_v = \int_{0}^{\infty} \omega F_v(\omega) \frac{\partial n_B(\omega, T_v)}{\partial T_v} d\omega \qquad (3)$$

Here, $N(\epsilon)$ and $F_v(\omega)$ are the density of states for electrons, strongly coupled (v = sc) and weakly coupled (v = wc) phonons obtained, for both SLG and tBLG, from tight-binding [5,6] and force constant [7] methods parameterized by first principles calculations, and $f_D$ ($n_B$) is the Fermi-Dirac (Bose-Einstein) distribution. The difference between weakly and strongly coupled phonons in graphene is addressed considering that the optical A'$_1$ (TO) and E$^2_g$ (LO) phonons in the vicinity of K and Γ points of the Brillouin zone are those with high electron-phonon coupling. In this way, in our model for SLG, only phonons with frequency close to (1350 ± 2) cm$^{-1}$ and (1580 ± 2) cm$^{-1}$ are considered in the calculation of the strongly coupled quantities, while the others modes are used for the weakly coupled ones. For tBLG, the additional low frequency and high density phonon modes close to ZA$_2$ (88 ± 2) cm$^{-1}$ at the Γ point are also considered as strongly coupled phonons. This mode is not present in the SLG because its polarization represents the out-of-plane breathing between two adjacent layers due to the weak van der Waals interaction.

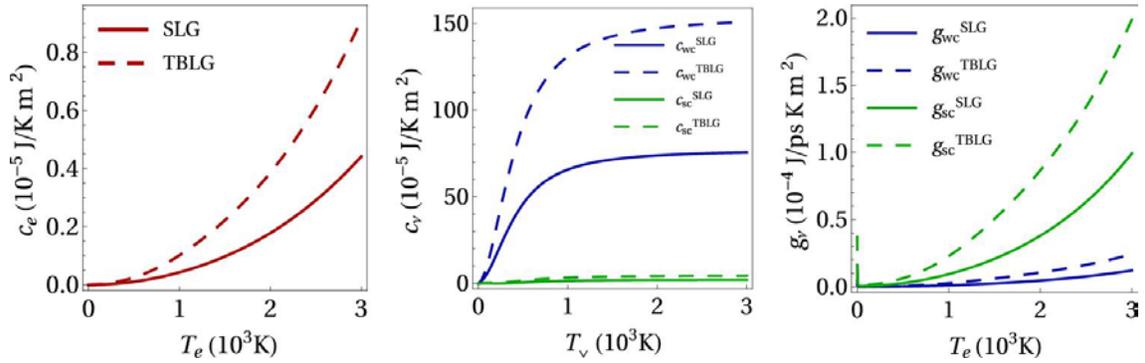

**Figure S-4** Temperature dependence of the specific heat for electron and phonons and the electron-phonon coupling factors for SLG and tBLG. The indexes *sc* and *wc* stand for strongly coupled and weakly coupled phonons, respectively.

The temperature dependent electron-phonon coupling factor can be calculated as in Ref. 8 by:

$$g_v = -\frac{\pi \hbar k_B \lambda_v \langle \omega^2 \rangle_v}{N(\epsilon_F)} \int_{-\infty}^{\infty} N^2(\epsilon) \frac{\partial f_D(\epsilon, T_e)}{\partial \epsilon} d\epsilon \qquad (4)$$

Figure S-4 shows the temperature dependence of the specific heat for electron and phonons and the electron-phonon coupling factors for SLG and tBLG. The parameters used are $\lambda_{sc} \langle \omega^2 \rangle_{sc}$ = 566 meV$^2$ and $\lambda_{wc} \langle \omega^2 \rangle_{wc}$ = 69 meV$^2$, that are the same values used for SLG [2].

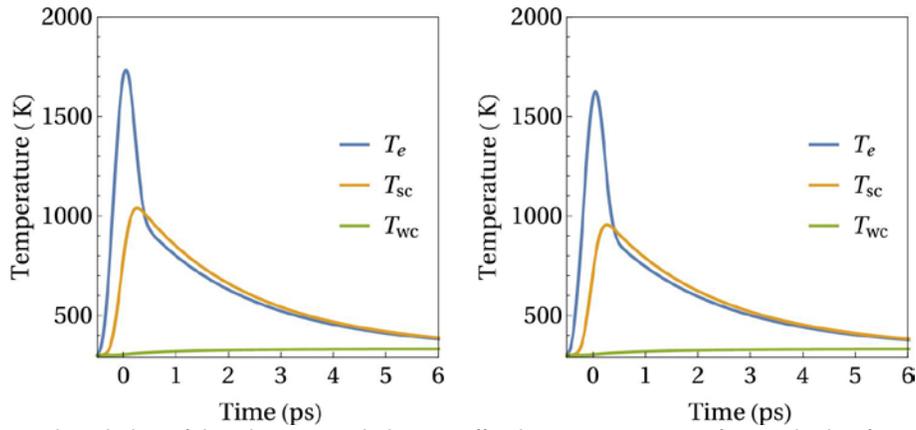

**Figure S-5** The temporal evolution of the electron and phonon effective temperatures after excitation for a laser fluence of 50 µJ cm$^{-2}$.

Figure S-5 shows the evolution of the effective temperatures for an incident fluence F = 50 µJ cm$^{-2}$ and a pulse duration of $\tau_p$ = 200 fs. To consider the difference in optical absorption for SLG and tBLG, we used β = 50 for the former and β = 25 for the latter. These values were chosen to compare with the experimental results.

## 4. Relaxation dynamics of tBLG stacks

Similar relaxation dynamics is measured in all the probed tBLG stacks presented in Figure 2 and 3 as long as the probe energy is tuned to match the energy of the three different bands we have discussed. The dynamics of tBLG are reported in Fig. S-6 as colored dots compared to that of the monolayer, displayed as black symbols. The probe energies are specified in the graphs, and correspond to the vHs (pink dots), the peak of the PA band (red symbols) and the low energy tail.

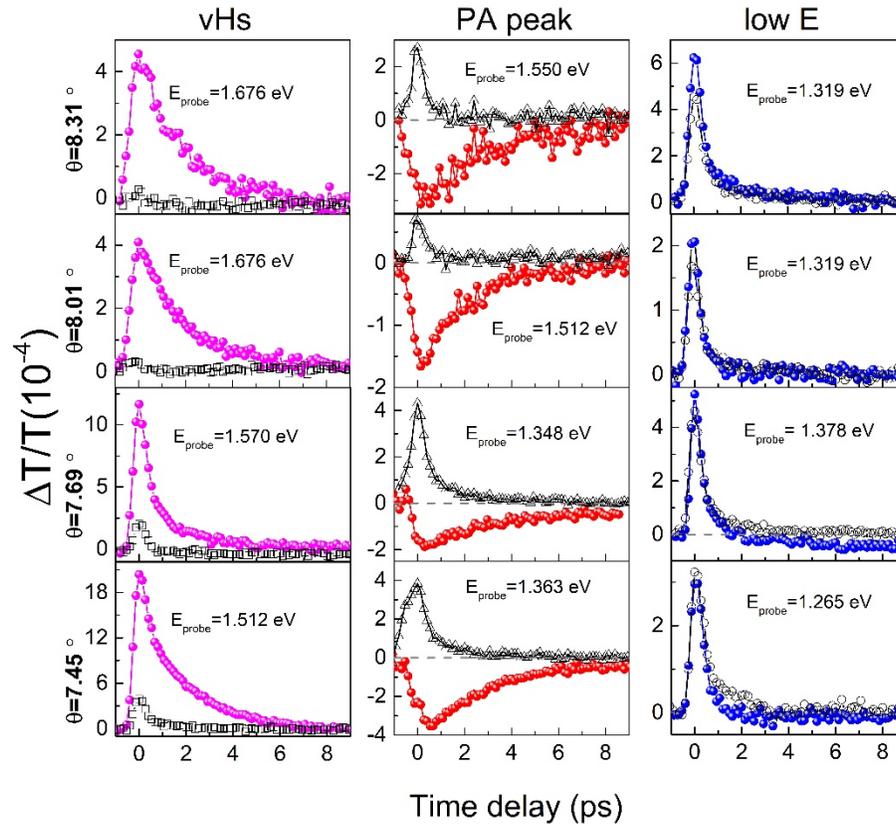

**Figure S-6** Transient transmissivity *ΔT/T* dynamics of tBLG (full colored dots) and SLG (open black symbols) for the stacks with θ = 8.31°, 8.01°, 7.69°, 7.45°, at the probe energies indicated in the graphs.